# Julia Augusta Taurinorum, an archaeoastronomical reload

**Amelia Carolina Sparavigna**
Politecnico di Torino

Discussion of an astronomical orientation of Turin as Julia Augusta Taurinorum, based on the results previously given in 2012 and available at arXiv:1206.6062, and here refined according to the local natural horizon, the refraction of atmosphere, and the use of software Stellarium. Besides the evaluation of a possible astronomical orientation, the use of the azimuth of Via Garibaldi, which is corresponding to the Roman decumanus, shows that the town is following the geometrical rules of Roman varatio.

Turin is a city located mainly on the left bank of the Po River, near the confluence with Dora River. A prehistoric settlement in this area existed since the third century BC. It was known as the village of Taurasia of a Celto-Ligurian people, the Taurini. It seems that the name of this population is coming from a Celtic word meaning "mountain", whereas local lores are referring to the bull (taurus). Heroically, Taurasia tried to impede the march of Hannibal when he was moving to attack Rome, coming from the Alps [1]. For three days the town resisted, but was eventually destroyed by Hannibal.

The origin of the modern city is, probably, in a castrum built by Julius Caesar during the Gallic Wars. Under Augustus, Turin became a Roman colony and its name, from Julia Taurinorum, turned into Julia Augusta Taurinorum, or simply, Augusta Taurinorum. Actually, the foundation of this colony is not mentioned in the Latin literature, and therefore information is coming only from historical and archaeological works [2]. The layout of the Roman colony is preserved in the oldest part of the city, which is known as the Quadrilatero Romano. In Ref. 3, it is told that the Roman town was determined by means of a "centuriation", that is, by a land limitation in the form of a rectangle of 770 m × 710 m, subdivided in 72 insulae (blocks). Francis John Haverfield, British historian and archaeologist, in his book on the planning of the ancient towns [4], appreciated Turin as the best preserved example of  Roman town planning.

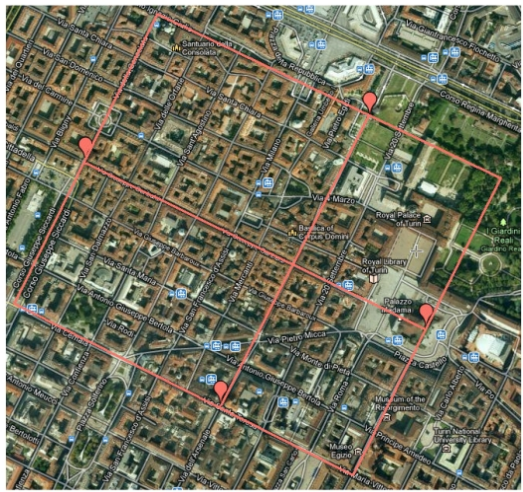

Fig.1 This is the Roman Torino from Acme Mapper. The places of the four gates are marked (two of them are still existing). The Decumanus Maximus is the inclined East-West line, coincident to the modern Via Garibaldi. Note that the blocks are, more or less, the old Roman insulae. The "umbilicus", the centre of the town, is at the crossing of Decumanus and Cardo.

We can see in the Figure 1 that the Roman layout structure, that of a grid of parallel and perpendicular streets, is perfectly maintained in the town that we see today, and not only in the oldest part of it. During the Middle Ages, the Roman layout had been slightly distorted by the renewal of buildings and the opening of streets inside the insulae. In 1736, the central part of the town was "rectified". Its main street was again a straight line as it was at the Roman time. The architect who planned the rectification was G.G. Plantery [5].

The "umbilicus", the center of the town, was at the crossing of the Decumanus Maximus and the Cardo Maximus, the two main streets of Roman colonies (Fig.1). Plantery restored perfectly the old Decumanus, turning it in today Via Garibaldi. The modern street traces perfectly the path of the Decumanus, starting from the East gate, now incorporated in Palazzo Madama, and ending at the West gate, the Porta Segusina. This gate was in the place where Via Garibaldi is crossed by Via della Consolata. The Porta Palatina, on the north side of the town is still well preserved and is the origin of the Cardo Maximus. Actually, under Via Garibaldi, it is running the old paved street and the greatest sewer of the Roman town. The sewer system was built in the first century AD, probably after the great fire of 69 AD, reported by Tacitus.

As previously told, the planning of the Roman town was performed by means of a centuriation [6], a method of land measurement and surveying. The centuriation is characterised by the regular grid traced using some surveyor's instruments. It seems that the foundation of a town followed a ritual [7], described by some Latin writers, during which an augur - a priest who practised divination - observing the flight of the birds and other omens, determined the ideal layout of the foundation. The gromatici, that is the surveyors, were deputed to transfer on the land the ideal plan devised by the augur. However, it is reasonable to imagine that, before the rituals of the foundation, the surveyors had already determined and prepared the best place for the town. Let us stress that some scholars are telling that the layout of the Roman colonies was determined by the local environment, "secundum naturam" [8,9]; other scholars insist on rituals determining the orientation of the town according to the rising of sun or moon, "secundum coelum". Also in Haverfield's book [4] we find mentioned the orientation to the sunrise of the decumanus of the Roman towns.

Ref.7 has investigated some ancient towns in Italy to evidence if they could have been determined according to an astronomical orientation linked to the sunrise. In [7], Torino is not discussed, moreover the azimuth of its decumanus is given with a wrong angle. It is given as an angle of 30 degrees, but this is not the angle of Torino that we can measure on the satellite maps (see Fig.1): the angle that the decumanus is forming with the cardinal East-West direction is of about 26 or 27 degrees (clockwise). The value depends of the satellite images that we use. Even considering that the sun has an apparent size of ½ degree, the angle given in Ref.7 is wrong. In [10], I considered the angle of the decumanus (I used a value of 25.8°), and determined two possible days, about which we could imagine the foundation of Julia Augusta Taurinorum, astronomically aligned to the sunrise. The days are 10 November and 30 January. In [11], I discussed possible festivals about these days, in particular the Kalends of February (in [12] the reader can find also a discussion of the work of Carlo Promis, about the archaeology of Turin). My work on the birthday of Torino was reported in 2016 by [13].

In [10], I have not considered the presence of the hills, east of the town, the refraction of the atmosphere, and the effect coming from small changes of the tilt of Earth's axis. Therefore, it seems necessary to me – because of news circulated in local TV on 26 December 2018, and because I want to stress, defend and reload the measurements I did in 2012 - to discuss these points, with the same aim of [10], which is that of using an approach that any person can repeat. For this reason, I follow an approach based on free software.

Let me stress that the aim of my works concerning archaeoastronomy is that of showing how people can easily perform some archaeoastronomical analyses. Of course, these are analyses that can be refined, but some relevant results can be obtained with my approach. For instance, I used it for Aosta [14], finding the results given in [15], and determined by direct observations.

**Natural horizon**

In any archaeoastronomical analysis we have to consider that the natural horizon can be quite different from the astronomical horizon which is usually given in calculations and software. In the case of Torino, it is so. Therefore, let us start by considering the presence of the hills, east of Torino (Figure 2). We can use Google Earth and draw a line along the decumanus. Let us prolong it on the hills. For the line corresponding to the decumanus (Via Garibaldi) and its prolongation, using the corresponding elevation profile given by Google Earth, we can obtain that the sun must reach an altitude of about 2.2 degrees on the astronomical horizon for being visible from the umbilicus of the town. Guido Cossard, author of book [16], confirmed via e-mail the value of about 2.2°. Let us note that we have to be careful in calculating this angle. For instance, the value of 2.2° is given by peak A in the Figure 2. Peak B is higher but it is quite distant from the umbilicus. Therefore it is giving a smaller angle. In the Figure 3, a drawing is illustrating this fact, when we have two peaks in the elevation profile. It is easy to generalize to more peaks.

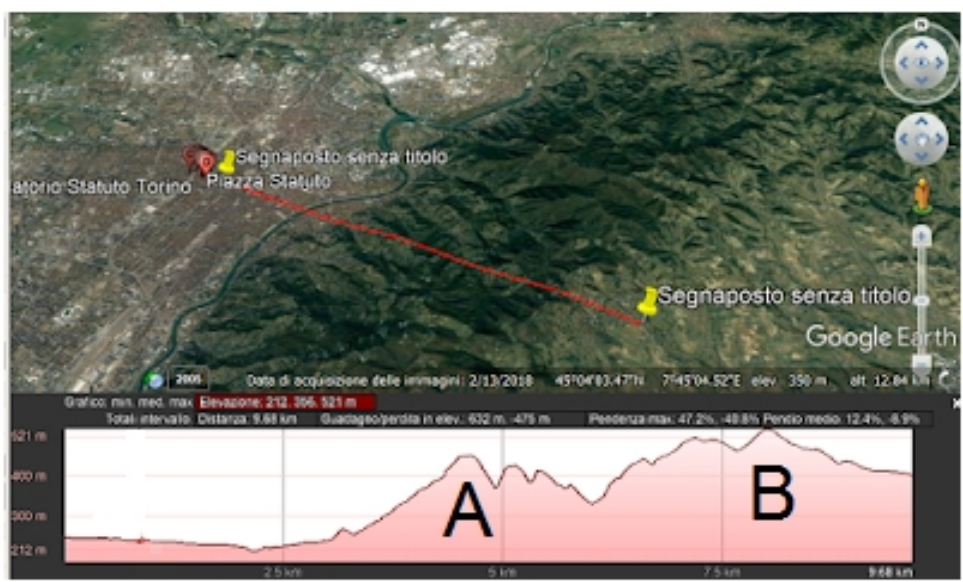

Figure 2. Let us use Google Earth and draw a line along the decumanus, and prolong it over the hills. In the image, you can see the elevation profile corresponding to the line. The sun must reach an altitude on the astronomical horizon of about 2.2 degrees to be visible from the umbilicus of the town. We have to be careful in calculating this angle. For instance, the value of 2.2° is given by peak A. Peak B is higher but it is so distant from the umbilicus, that it is giving a smaller angle (see please the following figure).

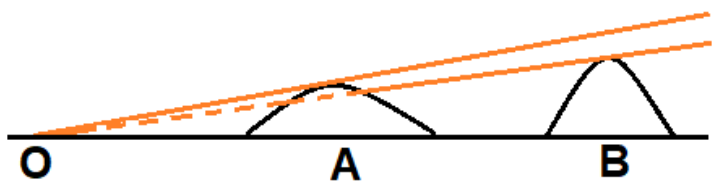

Figure 3. Illustration of the different angles for the altitude of the sun.

Instead of using the elevation profile, we can use the Google Maps. In the Figure 4, we can see a map that we can obtain of Torino and the nearby hills. From this map, a refined value of 2°14' can be obtained.

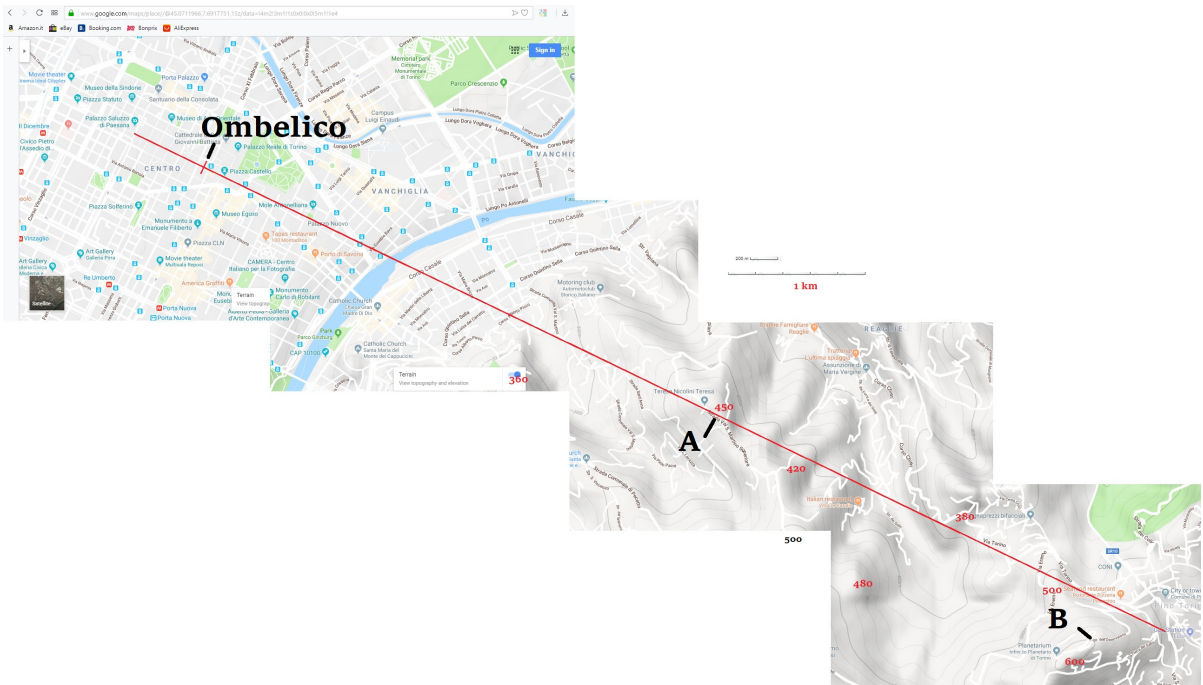

Figure 4. Thanks to Google Maps we can refine calculations.

Actually, we have also to consider the effect of the atmospheric refraction. It can be estimated to be of 18' (as given by Wikipedia). Therefore the altitude of the centre of the sun, above the astronomical horizon, must be (2°14'- 0°18'), to be observable from the umbilicus of Torino. However, we could imagine that it is the upper limb of the sun that was observed by the augur. So a further reduction of about 16' can be assumed. Therefore the altitude of the sun that we can consider is of about 1°40'.

**Software**
Having properly considered the natural horizon, we can use the altitude of the sun that we have determined in an astronomical software such as SunCalc.org.

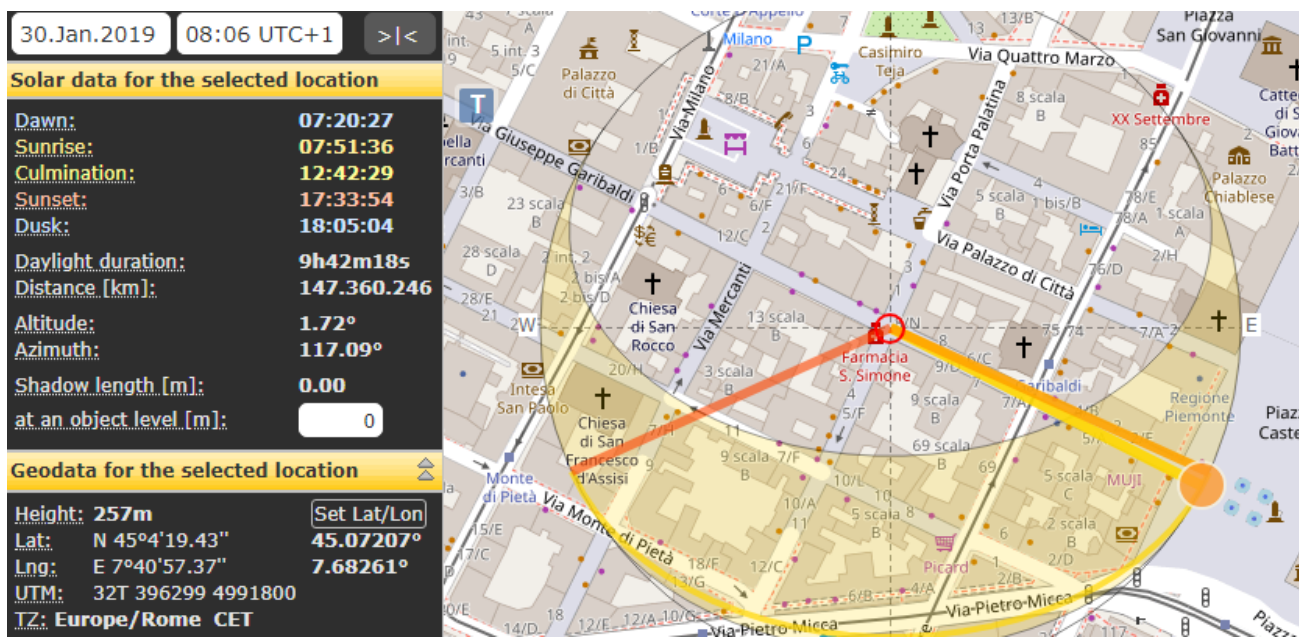

Figure 3. Screenshot of SunCalc.org, excellent software for astronomical analysis.

Let us consider this information for the day of 30 January, one of the days proposed in 2012 in [10]. Let us use SunCalc.org to find alignments along the direction of the rising sun on the natural horizon (altitude 1°40', that is, 1.70°). The result is given in the Figure 5; we can see the direction of the sunrise on the astronomical horizon represented by the orange line. We have a good agreement of the yellow line and the decumanus. The value of the azimuth is 117.1° (value from true North). We can also determine the uncertainty of the measure. The uncertainty of the azimuth obtained by means of SunCalc.org is 0.45 degree. Therefore, using SunCalc.org, the sunrise azimuth on the natural horizon is of 117.1° plus/minus 0.45°. We can repeat the analysis for 10 November, finding the same result as in the Figure 5. Actually, using software SunCalc.org, the natural horizon, the refraction of atmosphere, and observing the first limb of the sun, we have the same results as in [10].

Let us stress that, from an astronomical point of view, there is no reason to prefer 10 November or 30 January. Of course, we have to remember that a Roman town could have been planned just according to the local natural physical features of the area, and that the observed astronomical orientation could have no link to the day of the foundation.

**The azimuth of the Decumanus, used in Stellarium**
As we have previously told, Via Garibaldi is the perfect restoration of the Roman Decumanus. This is confirmed by a recent archaeological survey. This survey has produced a map available in the Geoportale del Comune di Torino, Visualizzatore SIT, webGis del Geoportale, in the category "Beni vincolati - Siti protetti – Archeologici". This map is given in the Figure 6.

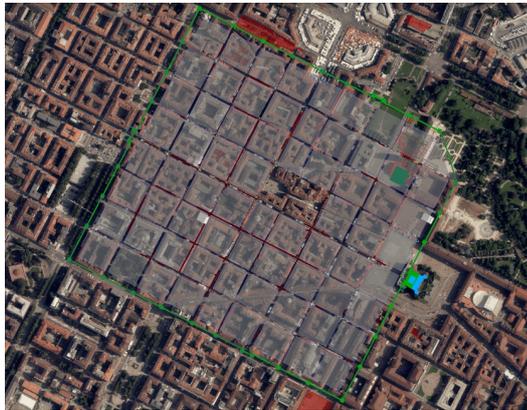

Figure 6. Many thanks to the Geoportale del Comune di Torino. The map is here used for research and study.

The map is given on a orthorectified satellite image. The angle the decumanus is forming with the vertical axis is 117.60°. that is 117°36'. This value is not so different from that given above. However, a difference exists, so let us try to use it. Moreover, another problem exists, and it is linked to the fact that, from the Roman time, the tilt of the Earth's axis is changed a little. So, it would be better to use a software, which is not limited to the present time. This software can be Stellarium. Therefore, fixing the altitude of the sun to the value of 1°40', that is the value that we have previously used, let us see what was the sunrise azimuth on 30 January of 9 BC, by means of Stellarium.

Why 9 BC? And also, why 117.60°? (I told that this is an angle the decumanus is forming with the vertical axis of the image. I am not telling that it is an azimuth, that is an angle from true North).
For year 9 BC, a reason exists, because this is the year mentioned by local TV on 26 December 2018. Also Ansa.it [17] and La Stampa of Torino, announced that two researchers had determined the date of the foundation of Torino, using the same approach as in [10], that is, to determine the day of the foundation by means of a comparison of the azimuth of the decumanus to that of the sunrise. Details of the work by Caranzano S. and Crosta M. is available in [18]. The authors tell that they have determined with *sufficient accuracy* the day and the year of foundation of the city: January 30, 9 BC. The reason for the accuracy is that the *astronomical date* is the same of the date of the inauguration of Ara Pacis in Rome by Augustus (inauguration of the festival of the Peace). Let me stress that the authors, in [18], are considering as coincident the *astronomical Julian date* (30 January 9 BC) to the *historical date* of the inauguration of the Ara Pacis. This is a mistake.
They have also given a wrong value of the altitude of the sun, because they have considered the peak B of Figure 2.
Since the authors of [18] are stressing that they are using a value of the *azimuth*, obtained by means of a measurement made using GPS, let us assume the angle 117° 40' [18]. If we use the proper value of the altitude of the sun in software Stellarium, we find that (let me note that 9 BC is given as -8 in Stellarium. Year "0" corresponds to 1 BC):

-8 Jan 30    az 118°28' alt 1°40'
-8 Jan 31    az 118°04' alt 1°40'
-8 Feb 1     az 117°39' alt 1°40'
-8 Feb 2     az 117°14' alt 1°40'
-8 Feb 3     az 116°49' alt 1°40'

Let us note that it is the sunrise azimuth of 1 February which is corresponding to the angle of the decumanus, very close to that obtained by the map of the archaeological survey, and proposed in [18]. In [18], the authors are stressing that 117°40' is an azimuth.

From calculations, we see that we have an effect of the change of the ecliptic which is relevant, if we want to determine the specific day of the foundation of Torino. At the same time, it means that the announcement of [17] is wrong.

**Discussion, first part**
As told before, Caranzano S. and Crosta M. tell that they have determined the day and the year of the foundation of Augusta Taurinorum [18], because the astronomical date seems coincident to the date of a Roman festival (the possibility that a Roman town was founded in coincidence to a Roman festival was a proposal of Magli in [7], but Magli's proposal is highly questionable, as shown in [19]). The date that Caranzano and Crosta have found, using the wrong value of the altitude of the sun, is the Julian date of 30 January. They fixed the year, because this date coincides, in their opinion, to that of the inauguration of the Ara Pacis in Rome by Augustus. Actually, the date of the inauguration of Ara Pacis is a historical date, given in the Julian Calendar, that does not correspond to the Julian date.

The Julian Calendar is the Calendar proposed by Julius Caesar in 46 BC, to substitute the Numa Calendar. It started from January 1, 45 BC (historical date). For many years the calendar operated by adding a leap year on a cycle of three years instead of four years. In 8 BC, Augustus stopped the intercalation. In 8 AD, the Julian Calendar was in agreement to the astronomical time, and started operating on a cycle of four years. Of the Julian Calendar, I discussed in [20] (see references therein). In my opinion, which is also the opinion of other scholars, the Calendar of Julius Caesar started on January 2 (Julian date, 45 BC). For what concerns the historical date of 30 January, that of the festival of the Peace (Ara Pacis), year 9 BC, it had to correspond to 3 (or 2 February) 9 BC (Julian date). The difference, that is 3 or 2 of February, is because some scholars are supposing the Julian Calendar as starting on the new moon of January 45 BC, others the day before. In my option, the Calendar of Julian Caesar started on the new moon, and therefore historical 30 January 9 BC is the Julian date of 3 February. In any case, the coincidence of the Julian date and historical date mentioned in [18] is impossible.

The Julian date of February 1, 9 BC, determined by means of Stellarium, corresponds to the historical 28 or 29 January 9 BC, before the inauguration of the Ara Pacis by Augustus! To have a coincidence, we need to wait to 4 BC, postponing the foundation of the town to 4 BC.

**Discussion, second part**

There is another problem in the article [18]. In it, the authors stress that, instead of using satellite images, they used measurements obtained by GPS of the direction of Via Garibaldi. In fact, a measurement of its azimuth, made by means of GPS or by means of a theodolite, must give the same result. In [21], we find a value of the measured azimuth of Via Garibaldi of 116.379° +/– 0.002°. A quite different value indeed.

In any case, let us use value [21], 116°23', in software Stellarium and maintain the same altitude of the sun previously used. Let us also maintain the same year. We obtain:

-8 Feb 3    az 116°50'  alt 1°40'
-8 Feb 4    az 116°23'  alt 1°40'
-8 Feb 5    az 115°58'  alt 1°40'

Let us change a little, of 10', the altitude of the sun. That is, we consider the value given by Cossard, obtained by the use of a clinometer We have:

-8 Feb 4    az 116°34'  alt 1°50'
-8 Feb 5    az 116°08'  alt 1°50'

In the case of an altitude of 2°, we have: -8 Feb 5    az 116°18'  alt 2°

Then, the day could have been 4 or 5 February 9 BC, in a Julian date. In the historical calendar of that year, it was January 31, or 1 or 2 of February. Then, the closest festival is the first of February, that is the Kalends of February.

The azimuth given in [21], is quite different from the "azimuth" given in [18]. Why? The reason is the following. In [18], the authors are **not** using an azimuth.

Let us use the GPS coordinates given in [18], and calculate their mean values for the east and the west ends of Via Garibaldi. The *grid angle* that we obtain is: 117.345°. Actually, the GPS coordinates are coordinates in the cartographic UTM Gauss projection of the earth globe. The true North, that is the geographic one, is different from the *grid* cartographic North. Therefore we have to pass from a *grid angle* to an *azimuth*, using the *convergence angle*. For Torino, applying the formulas given in [22], we have that this angle is of 57', that is 0.945°. We have to subtract this angle, because Torino is west the meridian of reference of 32T UTM zone. We find the azimuth of 116.40°, which is close to azimuth 116.38° given in [21].

Actually, [21] is in [23] article. In this new version of their proposal, Caranzano and Crosta continue to compare the grid angle to the sunrise azimuth. They are telling "La misura di 117,68° con il GPS, d'altra parte si accorda con un azimut ortivo del 30 gennaio per altezze della collina prossime a 1,80°, considerando il bordo superiore del Sole". They continue to use a wrong value of the altitude of the sun, but different from that used in [18]. In [23], they are neglecting the azimuth [21]. So, let me stress once more, they are continuing to use a grid angle, instead of an azimuth. Caranzano and Crosta continue to ignore the difference between the Julian dates and the dates of the historical calendar. Let me stress that the dates obtained in [23] are astronomical dates, not those of the Roman calendar! The 30 January that Caranzano and Crosta are proposing again in the new version of their work [23] is a Julian date, not the date of the Festival of the Pax! In any case, in the framework of [23], the coincidence is impossible.

Now, in the previous calculations, that we have here made using [21], we found 4 or 5 February,

astronomical dates, corresponding to 31 January, 1 or 2 February (historical dates), for 9 BC. From 8 BC, these dates are 2, 3 or 4 of February. In 8 AD, the historical dates were coincident to the astronomical dates for sure. In any case, the date that we find using the true azimuth is always different from that of the Festival of the Pax (historical 30 January).

I can imagine someone telling: the difference that you find was due to bad weather ... And in fact, this is also told in [23]: "sospensione dovuta al mal tempo". That is, the Romans suspended the foundation due to bad weather. Since the difference is systematic, it means that in Turin, it was always bad weather... In any case, if we assume that we can change the date because of bad weather, the use of astronomical calculation as made in [18] is meaningless, that is, it has no meaning or significance.

**Conclusion:** In my opinion, in the case of historical periods, archaeoastronomy can give interesting results. We have astronomical software, such as CalSKY and Stellarium and many others, which are suitable for obtaining proper quantitative data for any possible comparison. However, let me stress that the orientation of buildings and settlements "secundum naturam" must always be considered, and that an orientation "secundum coelum" has not to be considered, "a priori", the most relevant one. In the case of Torino, its orientation according to the local environment is perfect. Actually, there is no reason to prefer an orientation "secundum coelum" and fix consequently the date of the foundation. Moreover, using the azimuth that we can find in [21], we can see that Torino is following the rule of the varatio, which is using the geometry based on rectangular triangles in the planning of Roman town. As previously told, planning of Roman towns was based on the centuriation, a grid of parallel and perpendicular streets, where the two main axes were the Decumanus and the Cardo. By means of the varatio, the angle the Decumanus was forming with the east-west direction was given by the ratio of the catheti of a rectangular triangle. In the case of Torino [24], the town planning seems based on the ratio of catheti 1:2, which is giving an azimuth of the decumanus (Via Garibaldi) of 116,565° from true north. This value is in good agreement with the measurement of the direction of this street.

**Acknowledgement**
Many thanks to Ambrogio Manzino, Politecnico di Torino, for the very useful discussions on grid angles and azimuths.